\documentclass[aps,prd,twocolumn,superscriptaddress,floatfix,tightenlines]{revtex4}
\pdfoutput=1
\usepackage{bbm}
\usepackage{mathrsfs}
\usepackage{epsfig,psfrag}
\usepackage{amsmath,amsfonts,amssymb}
\usepackage[usenames]{color}
\usepackage{bm}
\usepackage{graphicx}
\usepackage{epstopdf}

\newcommand{\be}{\begin{equation}}
\newcommand{\ee}{\end{equation}}

\definecolor{BrickRed}{cmyk}{0,0.89,0.94,0.28}
\definecolor{MidnightBlue}{cmyk}{0.98,0.13,0,0.43}
\definecolor{DarkGreen}{rgb}{0,0.7,0.1}

\begin{document}


\title{Quantum and thermal Casimir interaction between a sphere and a plate:\\
  Comparison of Drude and plasma models}

\author{Roya Zandi} 
\affiliation{Department of Physics and Astronomy,
  University of California, Riverside, California 92521, USA}

\author{Thorsten Emig}
 \affiliation{Institut f\"ur Theoretische Physik, Universit\"at zu K\"oln,
 Z\"ulpicher Strasse 77, 50937 K\"oln, Germany}
 \affiliation{Laboratoire de Physique Th\'eorique et Mod\`eles
 Statistiques, CNRS UMR 8626, Universit\'e Paris-Sud, 91405 Orsay,
 France}
 
\author{Umar Mohideen}
\affiliation{Department of Physics and Astronomy,
  University of California, Riverside, California 92521, USA}

\date{\today}

\begin{abstract}
  We calculate the Casimir interaction between a sphere and a plate,
  both described by the plasma model, the Drude model, or
  generalizations of the two models. We compare the results at both
  zero and finite temperatures. At asymptotically large separations we
  obtain analytical results for the interaction that reveal a
  non-universal, i.e., material dependent interaction for the plasma
  model. The latter result contains the asymptotic interaction for
  Drude metals and perfect reflectors as different but universal
  limiting cases. This observation is related to the screening of a
  static magnetic field by a London superconductor.  For small
  separations we find corrections to the proximity force approximation
  (PFA) that support correlations between geometry and material
  properties that are not captured by the Lifshitz theory. Our results
  at finite temperatures reveal for Drude metals a non-monotonic
  temperature dependence of the Casimir free energy and a negative
  entropy over a sizeable range of separations.
\end{abstract}


\maketitle

\section{Introduction}
The past decade has witnessed rapid progress in the precision of
Casimir force measurements \cite{Lamoreaux97, Mohideen98,
  Harris:2000a, Decca:2005b,Decca07,Klimchitskaya:2009vn}. The
measurement precision that is expected in the near future demands
accurate theoretical calculations of the Casimir force for the
geometries and materials used in experiments. While Casimir's original
calculation for perfect metal plates \cite{Casimir48-1} and Lifshitz's
formula for dielectric slabs \cite{Dzyaloshinskii61} only apply to
planar, parallel surfaces, recent measurements have set limits on
geometry induced corrections in the most frequently used sphere-plate
geometry \cite{Krause:2007a}.  The geometry dependence of Casimir
forces is intriguing as it can vary substantially with the shape and
relative orientation of the objects
\cite{Emig:2009kx,rahi09,Bordag:2009uq}. Material dependence in the
form of dissipation in metals has been experimentally confirmed to
have an effect on the Casimir force \cite{Decca07,Chen07}.  It is thus
important that the geometry and material dependence be carefully
investigated for the experimentally most important sphere-plate
configuration.

In order to compare the experimental results to theory, the Derjaguin
or proximity force approximation (PFA) \cite{Parsegian05} has commonly
been used. This approximation neglects the non-additivity of Casimir
forces by estimating the interaction between curved surfaces in terms
of the planar surface interaction between infinitesimal and parallel surface
elements.  Its validity is hence limited to the singular limit of
vanishingly small separations between the surfaces. A systematic
extension to larger separations is not possible within such
approximations. 

The first exact computation of the Casimir interaction energy for a
perfectly reflecting sphere and plate was presented in
Ref.~\cite{Emig:2008ee}. Recently, corrections that come from using
the plasma or Drude model were computed at zero temperature
\cite{Canaguier09} and at $T=300^\circ K$
\cite{Canaguier-Durand:2010fk}. Other open geometries with curvature
such as a cylinder above a plate have been studied for perfect metals
\cite{Emig06}.  Corrections to the PFA in the case of perfect metals
for a cylinder above a plate and a sphere above a plate have been
obtained using path integral approaches \cite{Bordag:2006b,Bordag:kx}
and for scalar fields employing a world line formalism
\cite{Gies:2006b}.

Here we show that Casimir forces reveal a rich interplay between
geometry (radius of the sphere and object's separation), optical
properties of metals and thermal fluctuations.  We study this in
detail by calculating the Casimir interaction for different sphere
radii and separations using the (i) the Drude model, (ii) a
generalized Drude model, (iii) the plasma model and (iv) a generalized
plasma model at different temperatures.  The study of these combined
effects is of utmost importance since Casimir force measurements
continue to be carried out using this geometry and an increasing
accuracy is expected.  Hence, the experimental findings will begin to
show sensitivity to the material and temperature effects, which we
take into account here.  Furthermore, the unabated controversy whether
the plasma or the Drude model is more appropriate for describing the
optical properties of metals in Casimir calculations compels us to
provide results for both models so that experimentalists can build on
them when studying this problem further. The plasma model is a
high-frequency description of the optical properties and the
divergence $\sim 1/\omega^2$ of its dielectric function for small
$\omega$ is unphysical for metals.  The Drude model provides a proper
low-frequency description for metals with a $1/\omega$ divergence of
the dielectric function for small $\omega$. At large frequencies, both
models become identical.

Below, we supply numerical results for the Casimir interaction at
arbitrary separations as well as analytic formulas for the asymptotic
interaction at large separations. Depending on the model under
consideration, the asymptotic results show universal or non-universal
(i.e., material-independent or -dependent) behavior, a feature which
is not present for the simple case of two parallel metal plates and
hence results from the interplay of finite object sizes and material
properties.

\section{General expression for the interaction}
\label{sec:formulas}

To calculate the interaction of a metallic sphere of radius $R$ and a
metallic plate with a separation $d$ between the center of the sphere
and the plate, we employ a scattering approach for Casimir
interactions, which is described in detail in Ref.~\cite{rahi09}.  The
Casimir free energy of this system at temperature $T$ is given by
\begin{equation}
  \label{eq:energy_general}
  E = k_B T {\sum_{n=0}^{\infty}}^\prime \log \det [1-Y(\kappa_n)] \, ,
\end{equation}
with Matsubara wave numbers $\kappa_n=2\pi n k_B T/\hbar c$.  The
primed sum indicates that the contribution for $n = 0$
is to be weighted by a factor of $1/2$.  At zero temperature the sum
is replaced by an integral along the imaginary frequency axis \cite{rahi09},
\begin{equation}
  \label{eq:energy_general_T0}
  E = \frac{\hbar c}{2\pi} \int_0^\infty d\kappa \log \det [1-Y(\kappa)] \, ,
\end{equation}
where the matrix $Y$ is given by the product
\begin{equation}
  \label{eq:Y-matrix}
  Y^{\alpha\beta}_{lml'm'} = T_{s,lm}^\alpha U_{lml'm}^{\alpha\beta} \delta_{mm'}
\end{equation}
of the T-operator $T_s$ of the sphere and an operator $U$ that
describes the propagation of waves between the plate and the sphere
and the scattering of them at the plate (see below). We represent
these operators in a vector basis of spherical waves, where $\alpha$,
$\beta$=E, M denote electric or magnetic multipoles and $l$, $m$ label
the spherical waves.  For a sphere of radius $R$ with uniform
permittivity $\epsilon(\omega)$ and permeability $\mu(\omega)$ the
T-matrix elements for M-multipoles are given by
\begin{widetext}
\begin{equation}
\label{eq:t-matrix-elem-sphere}
  T^{\textsc{m}}_{s,lm}= \frac{\pi}{2} \frac{\eta I_{l+{1\over 2}}(\kappa R)
\left[I_{l+{1\over 2}}(n\kappa R)+2n\kappa RI'_{l+{1\over 2}}(n\kappa R)\right] - n I_{l+{1\over 2}}(n\kappa R)
\left[I_{l+{1\over 2}}(\kappa R)+2\kappa R I'_{l+{1\over 2}}(\kappa R)\right]}
{\eta K_{l+{1\over 2}}(\kappa R)
\left[I_{l+{1\over 2}}(n\kappa R)+2n\kappa RI'_{l+{1\over 2}}(n\kappa R)\right] - n I_{l+{1\over 2}}(n\kappa R)
\left[K_{l+{1\over 2}}(\kappa R)+2\kappa R K'_{l+{1\over 2}}(\kappa R)\right]} \, ,
\end{equation}
\end{widetext}
with $n=\sqrt{\epsilon(i\kappa)\mu(i\kappa)}$,
$\eta=\sqrt{\epsilon(i\kappa)/\mu(i\kappa)}$.  The T-matrix elements
for E-multipoles, $T^{\textsc{e}}_{s,lm}$, are obtained from
Eq.~(\ref{eq:t-matrix-elem-sphere}) by interchanging $\epsilon$ and
$\mu$ and by changing the overall sign. By taking
$\epsilon(i\kappa)\to\infty$ at an arbitrarily fixed $\mu(i\kappa)$ in
Eq.~(\ref{eq:t-matrix-elem-sphere}), the limit of a perfectly
reflecting sphere and plate is obtained. Then the matrix elements
become independent of $\mu$,
\begin{subequations}
\label{eq:t-matrix-elem-cond-sphere}
\begin{align}
  \label{eq:t-matrix-elem-cond-sphere-m}
   T^{\textsc{m}}_{s,lm}&=  \frac{\pi}{2} \frac{I_{l+{1\over 2}}(\kappa R)}{K_{l+{1\over 2}}(\kappa R)}
\\
  \label{eq:t-matrix-elem-cond-sphere-e}
T^{\textsc{e}}_{s,lm}&= - \frac{\pi}{2} \frac{I_{l+{1\over 2}}(\kappa R)+2\kappa RI'_{l+{1\over 2}}(\kappa R)}
{K_{l+{1\over 2}}(\kappa R)+2\kappa RK'_{l+{1\over 2}}(\kappa R)} \, .
\end{align}
\end{subequations}
These matrix elements scale for small $\kappa$ as $\kappa^{2l+1}$.  It
is interesting to compare this behavior to the scaling of the general
matrix elements of Eq.~(\ref{eq:t-matrix-elem-sphere}) for the
dielectric functions of the Drude and plasma model. For both models
the $T^{\textsc{e}}_{s,lm}\sim\kappa^{2l+1}$ behavior is unchanged for
 E-multipoles. The coefficients become material (plasma frequency)
dependent for the plasma model but retain the universal values of a
prefect reflector for the Drude model.  However, for M-multipoles only
the plasma model shows this universal behavior while the Drude model
yields a different scaling $T^{\textsc{m}}_{s,lm}\sim\kappa^{2l+2}$ with 
non-universal conductivity dependent coefficients. 

The operator $U$ can also be expressed in a spherical wave basis.  It
describes the propagation of waves from the sphere to the plate, a
reflection at the plate and the propagation back to the sphere.  The
reflection of waves at a dielectric plane is described most easily in
a plane wave basis with in-plane wave vector $k_\|$. The T-matrix
elements of the plane are then given by the usual Fresnel
coefficients. The conversion from plane to spherical waves and
simultaneous translation from the sphere to the plane is obtained by
multiplying the plane's T-matrix from left and right by a matrix
$D^{\alpha\beta}_{lm}(k_\|)$.  After defining $z=k_\|/\kappa$, the
matrix multiplication runs over the continuous variable $z$ and the
elements of the operator $U$ can be written as
\begin{equation}
  \label{eq:U-matrix}
  U_{lml'm}^{\alpha\beta}\! = \!\!\!\int_0^\infty \!\!\frac{z dz}{4\pi} \frac{e^{-2d\kappa
\sqrt{1+z^2}}}{\sqrt{1+z^2}} \!\!\sum_\gamma \!D^{\alpha\gamma}_{lm}(z) T_p^\gamma(\kappa,z)
{D_{l'm}^{\beta\gamma}}^*(z) \, ,
\end{equation}
where the plate's diagonal T-matrix, $T_p^\gamma(\kappa,z)$, for
polarization $\gamma$ are given by
\begin{subequations}
  \label{eq:T_matrix_plane}
\begin{align}
  T_p^{\textsc{M}} &= -\frac{\mu(i\kappa)\sqrt{1+z^2}-
\sqrt{\epsilon(i\kappa)\mu(i\kappa)+z^2}}{\mu(i\kappa)\sqrt{1+z^2}+
\sqrt{\epsilon(i\kappa)\mu(i\kappa)+z^2}}\, ,\\
 T_p^{\textsc{E}} &= \frac{\epsilon(i\kappa)\sqrt{1+z^2}-
\sqrt{\epsilon(i\kappa)\mu(i\kappa)+z^2}}{\epsilon(i\kappa)\sqrt{1+z^2}+
\sqrt{\epsilon(i\kappa)\mu(i\kappa)+z^2}} \, .
\end{align}
\end{subequations}
The exponential factor in Eq.~(\ref{eq:U-matrix}) describes the translation
from the sphere to the plane and back by a total distance $2d$ in the
plane wave basis. The elements of the matrix that converts between
plane and spherical waves are given by
\begin{subequations}
\begin{align}
  \label{eq:D-matrix-elements}
D^{\textsc{MM}}_{lm} &= D^{\textsc{EE}}_{lm} = \sqrt{\frac{4\pi(2l+1)(l-m)!}{l(l+1)(l+m)!}} 
z {P_l^m}'\left(\sqrt{1+z^2}\right) \, ,\\ 
D^{\textsc{EM}}_{lm} &= -D^{\textsc{ME}}_{lm} = \sqrt{\frac{4\pi(2l+1)(l-m)!}{l(l+1)(l+m)!}} 
\frac{im}{z} P_l^m\left(\sqrt{1+z^2}\right) ,
\end{align}
\end{subequations}
where $P_l^m$ is the associated Legendre polynomial of order $l$, $m$.
These elements have the following symmetries under complex
conjugation,
\begin{subequations}
\begin{align}
  \label{eq:D_symm}
  {D^{\alpha\alpha}_{lm}}^* &= (-1)^m D^{\alpha\alpha}_{lm} \,\\
  {D^{\textsc{ME}}_{lm}}^* &= (-1)^{m+1} D^{\textsc{ME}}_{lm} \, . 
\end{align}
\end{subequations}
In what follows we employ Eqs.~(\ref{eq:energy_general}),
(\ref{eq:energy_general_T0}) to obtain the Casimir interaction for
perfectly reflecting bodies and also for metals described by the
plasma and Drude model at zero and finite temperatures.

\section{Large distance interaction at $T=0$ }

In this section we consider the zero temperature Casimir interaction
at large separations for different dielectric functions.

\subsection{Perfect reflector}

In the limit of perfect reflectivity of the plate, one finds from
Eq.~(\ref{eq:T_matrix_plane}) with $\epsilon\to\infty$ the simple
T-matrix elements $T_p^{\textsc{M}}= T_p^{\textsc{E}}=1$.  With this
simplification, the integration over $z$ in Eq.~\eqref{eq:U-matrix}
can be performed analytically. We find for the elements of $U$ the
same result that was obtained before, using the method of images
\cite{Emig:2008ee},
\begin{widetext}
\begin{eqnarray}
  \label{eq:U-perfect-metal}
  U^{\textsc{MM}}_{lml'm}&=&U^{\textsc{EE}}_{lml'm}=(-1)^{l+l'+1} \sum_{l''=|l-l'|}^{l+l'} 
\frac{(-1)^{l''}}{2} 
\left[l(l+1)+l'(l'+1)-l''(l''+1)\right] \sqrt{\frac{(2l+1)(2l'+1)}{l(l+1)l'(l'+1)}}
(2l''+1)\nonumber\\
&&\times\begin{pmatrix}l'&l&l''\\0&0&0\end{pmatrix}
\begin{pmatrix}l'&l&l''\\m&-m&0\end{pmatrix} 
\frac{K_{l''+1/2}(2\kappa d)}{\sqrt{\pi \kappa d}} \\
 U^{\textsc{ME}}_{lml'm}&=&-U^{\textsc{EM}}_{lml'm}=(-1)^{l+l'+1}
2i \kappa d m \sum_{l''=|l-l'|}^{l+l'} (-1)^{l''} 
\sqrt{\frac{(2l+1)(2l'+1)}{l(l+1)l'(l'+1)}}
(2l''+1)\nonumber\\
&&\times\begin{pmatrix}l'&l&l''\\0&0&0\end{pmatrix}
\begin{pmatrix}l'&l&l''\\m&-m&0\end{pmatrix} 
\frac{K_{l''+1/2}(2\kappa d)}{\sqrt{\pi \kappa d}} \, .
\end{eqnarray}
\end{widetext}
Using this result and the T-matrix elements of
Eq.~\eqref{eq:t-matrix-elem-cond-sphere} we obtain for the interaction
energy the large distance expansion
\begin{equation}
  \label{eq:E_perf_large_d}
  E = -\frac{\hbar c}{\pi} \left( \frac{9}{16} \frac{R^3}{d^4} +
    \frac{25}{32} \frac{R^5}{d^6} +{\cal O}(R^6/d^7) \right) 
\end{equation}
at zero temperature \cite{Emig:2008ee}.

\subsection{Plasma model}
\label{sec:plasma_exact}

We now assume that both the sphere and the plate are described by the
plasma model which on the imaginary frequency axis has the dielectric
function
\begin{equation}
  \label{eq:eps_plate}
  \epsilon_p(ic\kappa) = 1+ \left(\frac{2\pi}{\lambda_p\kappa}\right)^2 
\end{equation}
The plasma wavelength $\lambda_p$ is related to the plasma frequency
$\omega_p$ by $\lambda_p = 2 \pi c/\omega_p$.  Note that the plasma
model provides a high-frequency description of optical properties and
ignores dissipation. Hence it is not expected to capture the low
frequency response of a metal. To understand the physical meaning
of the results for the Casimir interaction presented below, it is interesting
to realize that the dielectric function of Eq.~\eqref{eq:eps_plate}
appears also in the wave equation for the magnetic field in a
superconductor when it is described by the London theory. The second
London equation and the Maxwell equations yield $\epsilon_p$ with
the penetration depth $\lambda_p=\sqrt{m_e c^2/(16\pi^3 n_s e^2)}$
for superfluid carriers of density $n_s$, charge $e$ and mass $m_e$. 

To obtain the large distance behavior of the Casimir energy, we need
to expand the T-matrices for small $\kappa$.  To this end, we set
$\kappa=u/d$ and expand the relevant expressions in powers of $1/d$.
The T-matrix elements of the sphere scale as $\kappa^{2l+1}$ for
$\kappa\to 0$ for both E and M polarizations.  In the case of the E
polarization the coefficients are universal and are given by the
perfect reflector result which corresponds to
\begin{equation}
  \label{eq:T_E_small_kappa}
  T^{\textsc{E}}_{s,lm} = \frac{l+1}{l} \frac{1}{(2l+1)!!(2l-1)!!} (u R/d)^{2l+1} + \ldots \, .
\end{equation}
However, for the M polarization the coefficients are not universal and
depend on  the plasma wave length as follows
\begin{equation}
  \label{eq:T_M_small_kappa}
  T^{\textsc{M}}_{s,lm} = \frac{I_{l+{3\over 2}}(2\pi
    R/\lambda_p)}{I_{l-{1\over 2}}(2\pi R/\lambda_p)} 
\frac{1}{(2l+1)!!(2l-1)!!} (u R/d)^{2l+1} + \ldots \, .
\end{equation}
In the limit of a small plasma wavelength, $\lambda_p \ll R$, the
elements of this matrix approach the perfect reflector limit with is
given by
\begin{equation}
  \label{eq:T_M_small_kappa_perf}
  T^{\textsc{M}}_{s,lm} = 
\frac{1}{(2l+1)!!(2l-1)!!} (u R/d)^{2l+1} + \ldots \, .
\end{equation}
For a large plasma wavelength, $\lambda_p\gg R$, the elements
are not universal and reduced by a factor $(R/\lambda_p)^2$ compared
to the perfect reflector limit,
\begin{equation}
  \label{eq:T_M_small_kappa_large_lp}
  T^{\textsc{M}}_{s,lm} = 
\frac{(2\pi R/\lambda_p)^2}{(2l+1)!!(2l+3)!!} (u R/d)^{2l+1} + \ldots \, .
\end{equation}
The latter result can be understood in terms of the London
superconductor interpretation of the plasma model. If the penetration
depth $\lambda_p$ becomes much larger than the radius, the sphere
becomes almost transparent for the magnetic field and the T-matrix
elements are reduced to small values $\sim R^{2l+3}/\lambda_p^2$.

The T-matrix elements of the plate with $\epsilon_p(ic\kappa)$ of
Eq.~\eqref{eq:eps_plate} depend also on the lateral wave vector ${\bf
  k}_\|$. To obtain the large distance expansion, we set $k_\|=v/d$
and expand the T-matrix for large $d$ with $z=k_\|/\kappa=v/u$ fixed.
This yields the expansion of the plate's T-matrix elements,
\begin{equation}
  \label{eq:T-plate_large_d}
\begin{split}
  T^M_p &= 1-\frac{\sqrt{z^2+1} u\lambda_p }{\pi  d}+\frac{ \left(z^2+1\right) u^2\lambda_p ^2}{2 \pi ^2 d^2}+{\cal O}\left(u^3\right) \\ 
 T^E_p &= 1-\frac{u \lambda_p }{\pi  d \sqrt{z^2+1}}
+\frac{u^2 \lambda_p ^2}{2 \pi ^2 d^2 \left(z^2+1\right)}+{\cal
  O}\left(u^3\right) \, .
\end{split}
\end{equation}
With this expansion, the integral over $z$ in Eq.~\eqref{eq:U-matrix}
can be performed analytically, and one obtains an expansion in $1/d$
of the matrix elements of $U$ which depend on $u$ and $\lambda_p/d$
only.  When we substitute the matrix elements of
Eqs.~\eqref{eq:T_E_small_kappa}, \eqref{eq:T_M_small_kappa}
and \eqref{eq:T-plate_large_d} into Eq.~\eqref{eq:Y-matrix} and expand
the energy of Eq.~\eqref{eq:energy_general} in powers of $1/d$, we
obtain the interaction to order $1/d^6$ by including $l=2$ partial
waves.  The result can be written as
\begin{widetext}
\begin{equation}
  \label{eq:energy-plasma}
    {\mathcal E} = -\frac{\hbar c}{\pi}\left[ f_4(\lambda_p/R) \frac{R^3}{d^4}
+f_5(\lambda_p/R)\frac{R^4}{d^5}  +f_6(\lambda_p/R)\frac{R^5}{d^6}
+{\cal O}(d^{-7}) \right] 
\end{equation}
with the functions
\begin{subequations}
\begin{align}
  \label{eq:f-fct}
f_4(z) & =\frac{9}{16} +\frac{9}{64\pi^2} z^2 -\frac{9}{32\pi} z
\coth\frac{2\pi}{z}\, ,\\
f_5(z) & = - \frac{13}{20\pi}z-\frac{21}{80\pi^3} z^3
+\frac{21}{40\pi^2} z^2 \coth \frac{2\pi}{z} \, , \\
f_6(z)&=\frac{1}{1792(2\coth(2\pi/z)-z/\pi)}
\left[\left( 2800 +\frac{2595}{\pi^4} z^4 
+\frac{10072}{\pi^2} z^2 \right) \coth(2\pi/z)\right.
\nonumber\\
&-\left.\frac{z/\pi}{\sinh^2(2\pi/z)}\left(-2100 -\frac{285}{\pi^4} z^4
-\frac{223}{\pi^2} z^2 + \left( 
3780+\frac{285}{\pi^4} z^4 +\frac{3763}{\pi^2}z^2 \right) \cosh(4\pi/z) \right.\right.\nonumber\\
&-\left.\left.\frac{1260}{\pi} z \coth(2\pi/z)
\right)
\right]
\end{align}
\end{subequations}
\end{widetext}
Note the coefficient $f_4$ of the leading term depends on $\lambda_p$
and hence is not universal. Only in the two limits $\lambda_p/R\to 0$
and $\lambda_p/R\to\infty$ the coefficient approaches the material
independent values $9/16$ and $3/8$, respectively. This behavior is
consistent with the two limiting forms of the sphere's T-matrix of
Eqs.~\eqref{eq:T_M_small_kappa_perf},
\eqref{eq:T_M_small_kappa_large_lp}. The limit $\lambda_p/R\to 0$
describes perfect reflection of electric and magnetic fields at
arbitrarily low frequencies and hence agrees with the result of
Eq.~\eqref{eq:E_perf_large_d} where for dipole fluctuations the E
polarization yields twice the contribution of the M polarization,
cf.~Eq.~\eqref{eq:T_E_small_kappa} and
Eq.~\eqref{eq:T_M_small_kappa_perf} for $l=1$. For $\lambda_p/R\to
\infty$ the coefficient $f_4$ is reduced by a factor $2/3$ since the M
polarization does not contribute to the leading term $\sim R^3/d^4$
due its suppression by $(R/\lambda_p)^2$,
cf.~Eq.~\eqref{eq:T_M_small_kappa_large_lp}.  Physically, the
non-universal behavior of $f_4$ can be understood when the objects are
considered as London superconductors.  For $\lambda_p/ R\to 0$ a
static magnetic field is perfectly screened and the objects become
perfect reflectors. If $\lambda_p/R \gg 1$, a static magnetic field
can penetrate the entire sphere and hence the M polarization does not
contribute to the Casimir energy. From this interpretation it follows
that normal metals, which can be penetrated by a static magnetic
field, should interact to leading order in $R/d$ only via E
polarizations leading to $f_4=3/8$. We shall reach the same conclusion
when we consider the Drude model below.  The coefficient of $R^4/d^5$
is always positive and varies between $(13/20\pi^2)(\lambda_p/R)$ for
$\lambda_p/R\to 0$ and $(3/10\pi^2)(\lambda_p/R)$ for $\lambda_p/R\to
\infty$. The coefficient of $R^5/d^6$ can be negative (for
$\lambda_p/R\to 0$) or positive. In Section\ref{sec:numerics}, we
compare the exact findings of this Section to our results from a
numerical evaluation of Eq.~\eqref{eq:energy_general_T0} over a wide
range of separations.

Finally, it is instructive to compare the above results to the
interaction between two parallel and infinite plates that are
described by the plasma model. In this case, the large distance
expansion applies to $d \gg \lambda_p$ and the leading term is given
by the universal perfect reflector result. The plasma wavelength
appears only in corrections to the leading term that can be expanded
in powers of $\lambda_p/d$. This universal behavior is a consequence
of the (unrealistic) assumption of an infinite lateral size of the
plates which removes any finite length scale of the object that could be
compared to $\lambda_p$. Hence, a finite penetration depth only yields
an increased effective separation which for $d \gg \lambda_p$
obviously approaches $d$, explaining the universal large-$d$ result.

\subsection{Drude model}
\label{sec:drude_exact}

The Drude model describes the low-frequency response of a metal
which depends on its dc conductivity $\sigma$. For large
frequencies it becomes identical to the plasma model with plasma
wavelength $\lambda_p$. On the imaginary frequency axis, the Drude
dielectric function is given by
\begin{equation}
  \label{eq:eps_Drude}
  \epsilon_D(ic\kappa)=1+\frac{(2\pi)^2}{(\lambda_p\kappa)^2+
\pi c \kappa/\sigma_p} \, ,
\end{equation}
The conductivity is associated with the length scale
$\lambda_\sigma=2\pi c/\sigma$.  At large distances $d$, we need to
consider the limit of small $\kappa$ at fixed $z=k_\|/\kappa$ for the
plate's T-matrix, which yields with $\kappa=u/d$
\begin{subequations}
  \label{eq:T-plate_Drude_large_d}
\begin{align}
  T^M_p &= 1- \sqrt{\frac{uc}{\pi d \sigma_p}}\sqrt{z^2+1}+\frac{uc}{2\pi d \sigma_p} \left(z^2+1\right)+{\cal O}\left(u^{3/2}\right) \\
 T^E_p &=  1-\sqrt{\frac{uc}{\pi d \sigma_p}}\frac{1}{\sqrt{z^2+1}}
+\frac{uc}{2\pi d \sigma_p}\frac{1}{z^2+1}+{\cal
  O}\left(u^{3/2}\right)  .
\end{align}
\end{subequations}
The approach of unity for both polarizations is a consequence of
keeping $k_\|/\kappa$ fixed in the limit $\kappa\to0$. This behavior
arises from the fact that the plates are infinitely extended so that
arbitrarily small $k_\|$ are allowed. The situation is different at
finite temperatures where one has to take $\kappa\to 0$ at fixed
$k_\|$ for the first term of the sum over Matsubara frequencies. In
the latter limit the magnetic contribution $T^M_p$ vanishes. 

For the sphere with the Drude dielectric function of
Eq.~\eqref{eq:eps_Drude} we obtain for the T-matrix elements with
$l=1$ the low frequency expansion
\begin{subequations}
\begin{align}
  \label{eq:T-sphere-Drude}
  T_{s,1m}^{M} &= \frac{4\pi}{45} \frac{R\sigma_p}{c}
(uR/d)^4 + \ldots \\
T_{s,1m}^{E} &= \frac{2}{3} (uR/d)^3 -\frac{1}{2\pi} \frac{c}{R\sigma_p} (uR/d)^4 +\ldots \, .
\end{align}
\end{subequations}
While the leading term of the E polarization agrees with the perfect
reflector result, the leading term of the M polarization is reduced by
a factor $R\kappa=uR/d$ compared to the perfect reflector case.
Therefore, one expects that only the E polarization contributes to the
leading term of the interaction at large distances.

With the above expansion of the T-matrix elements the integrations 
over $u$ and $z$ can be performed and from the dipole contributions
with $l=1$ we obtain for the energy the large distance expansion
\begin{equation}
\begin{split}
  \label{eq:E_Drude_both}
  E &= -\frac{\hbar c}{\pi} \left[
\frac{3}{8} \frac{R^3}{d^4} -\frac{77}{384} \frac{R^3}{\sqrt{2\sigma
    /c} \, d^{9/2}}\right.\\
&-\left. \frac{c R^3}{8\pi \sigma d^5}
+\frac{\pi}{20}\frac{\sigma}{c}\frac{R^5}{d^5}+{\cal O}(d^{-\frac{11}{2}})\right] \, .
\end{split}
\end{equation}
The leading term in Eq.~\eqref{eq:E_Drude_both} shows the universal
amplitude coming only from the E polarization as expected from the
form of the T-matrix elements. This result reproduces the
prediction of the plasma model in the limit where $\lambda_p \gg R$,
see the discussion below Eq.~\eqref{eq:f-fct}. This limit describes
the situation where a static magnetic field can fully penetrate the
sphere and hence describes a normal metal. The correlations between
material and shape become obvious when one compares the above result
to the interaction between two parallel and infinite plates that are
described by the Drude model. For this geometry the large distance
expansion applies to $d\gg c/\sigma$. The leading term of this
expansion is identical to the prefect reflector result, as for the
plasma model. The dc conductivity appears only in corrections to the
leading term that can be expanded in integer powers of $\sqrt{c/\sigma
  d}$. Since the frequently used PFA for the sphere-plate geometry is
based on the two-plate energy, it would predict at sufficiently large
$d$ for both the plasma and the Drude model the perfect reflector
result which has equal contributions from E and M
polarization. However, it is known that the PFA does not apply to
large distances. It should be noted that the result of
Eq.~\eqref{eq:E_Drude_both} cannot be applied to an arbitrarily large
dc conductivity $\sigma$ since then the term $\sim R^5$, which comes
from the M polarization of the sphere, diverges.  The condition for
the validity of Eq.~\eqref{eq:E_Drude_both} can be written as $d \gg
R$, $\lambda_\sigma$, $R^2/\lambda_\sigma$. Below we shall study the validity
range of this expansion further by comparing it to numerical results.

\section{High temperature limit}
\label{sec:kappa=0}

In this section, we study the high temperature limit of the
sphere-plate interaction for the plasma and Drude model.  In this
case, the interaction is given by the first term of the Matsubara sum
of Eq.~\eqref{eq:energy_general}.  Hence we have to compute the
matrix elements of $Y(\kappa_0=0)$. This zero-frequency result
will turn out to be also useful when computing the Casimir energy
at zero and finite temperatures below since the limit $\kappa\to 0$
is numerically unstable due to the divergence of certain Bessel
functions. 

\subsection{Plasma model}

Here we have to consider the limit $\kappa\to 0$ at {\it fixed} $k_\|$
of the T-matrix of the plate since we are interested in arbitrary
separations $d$. In this limit the T-matrix elements are given by
\begin{equation}
 \label{eq:T_plate_plasma_small_kappa}
  T_p^M =  - \frac{|k_\||-\sqrt{4\pi^2/\lambda_p^2+k_\|^2}}
{|k_\||+\sqrt{4\pi^2/\lambda_p^2+k_\|^2}} \, ,\quad
 T_p^E  = 1 \, .
\end{equation}
The elements for the M polarization are non-universal and vary between
$1$ for $\lambda_p\to 0$ (perfect reflector) and $0$ for
$\lambda_p\to\infty$. The latter limit can be interpreted as
a London superconductor with diverging penetration depth such
that the plate is transparent to a static magnetic field.

For $\lambda_p\to 0$ the U-matrix of Eq.~\eqref{eq:U-matrix} can be
obtained for $\kappa\to 0$ from Eq.~\eqref{eq:U-perfect-metal}. To
obtain the U-matrix for $\kappa\to 0$ at non-zero but small $\lambda_p
\ll d$, we set $k_\|=v/d$ and expand $T_p^M$ of
Eq.~(\ref{eq:T_plate_plasma_small_kappa}) in $\lambda_p/d$ so that the
integral of Eq.~(\ref{eq:U-matrix}) can be performed
analytically. Since we are interested in the limit $\kappa\to 0$, we
only need the conversion matrix elements $D^{\alpha\gamma}_{lm}(z)$
for large arguments $z$. At large $z$ the associated Legendre
poynomials $P_{lm}(z)$ assume the limiting form $(-i)^m(2l-1)!!
z^l/(l-m)!$. Using the integral $\int_0^\infty e^{-2v} v^n dv =
n!/2^{n+1}$ we obtain to leading order for small $\kappa$ the matrix
elements
\begin{widetext}
\begin{subequations}
  \label{eq:U_small_kappa}
\begin{align}
  U^{EE}_{lml'm} &= \sqrt{\frac{l(2l+1)l'(2l'+1)}{(l+1)(l'+1)}}
\frac{(2l-1)!!(2l'-1)!!(l+l')!}
{\sqrt{(l+m)!(l-m)!(l'+m)!(l'-m)!}} \frac{1}{(2\kappa d)^{l+l'+1}} \\
U^{MM}_{lml'm}&= U^{EE}_{lml'm} \left[ 1- \frac{l+l'+1}{2\pi}
  \frac{\lambda_p}{d} + {\cal O} \left( (\lambda_p/d)^2\right) \right]
\, .
\end{align}
\end{subequations}
\end{widetext}
The matrix elements $ U^{EM}_{lml'm}$, $ U^{ME}_{lml'm}$ scale for
small $\kappa$ as $(\kappa d)^{-l-l'}$ and hence can be ignored.  The
T-matrix elements of the sphere for $\kappa\to 0$ are given by
Eqs.~\eqref{eq:T_E_small_kappa},~\eqref{eq:T_M_small_kappa} and hence
scale as $\kappa^{2l+1}$. The low-$\kappa$ scaling of the matrix
elements of $U$ and $T_s$ shows that the elements of the matrix $Y$
scale as $Y^{EE}_{lml'm} \sim Y^{MM}_{lml'm} \sim \kappa^{l-l'}$ and
$Y^{EM}_{lml'm} \sim Y^{ME}_{lml'm} \sim \kappa^{l-l'+1}$.  Hence for
$\kappa\to 0$ the coupling of E and M polarization does not contribute
to the energy.  We set again $\kappa=u/d$ and introduce the rescaled
matrix $\tilde Y$ with elements $\tilde Y_{lml'm}= u^{-l} Y_{lml'm}
u^{l'}$ so that divergences for $u\to 0$ are removed and in that limit
all elements of $\tilde Y^{MM}$ and $\tilde Y^{EE}$ assume non-zero
finite values that depend on $R/d$ and $\lambda_p/d$, and all elements
of $\tilde Y^{ME}$ and $\tilde Y^{EM}$ vanish. The rescaling does not
change the determinant of Eq.~\eqref{eq:energy_general} so that $\det
Y = \det \tilde Y$. In the high temperature limit the energy can then
be written as
\begin{equation}
  \label{eq:energy_high_T_plasma}
  E = \frac{k_B T}{2}   \log \det 
\begin{pmatrix}
1-\tilde Y^{MM}(u\to 0) & 0\\
0 & 1-\tilde Y^{EE}(u\to 0)
\end{pmatrix} \, ,
\end{equation}
where the matrix elements of $\tilde Y$ are given by
Eqs.~\eqref{eq:Y-matrix}, \eqref{eq:T_E_small_kappa},
\eqref{eq:T_M_small_kappa} and \eqref{eq:U_small_kappa}. 
By truncating the matrix $\tilde Y$ at lowest order $l=1$ we get
the high-temperature free energy 
\begin{equation}
  \label{eq:2}
  \begin{split}
    E & = -k_B T \left\{\left[ \frac{3}{8} +\frac{3}{32\pi^2}
      \frac{\lambda_p^2}{R^2} -\frac{3}{16\pi}\frac{\lambda_p}{R}
      \coth\left(2\pi \frac{R}{\lambda_p}\right) \right]
    \frac{R^3}{d^3} \right.\\
& \left. - \left[\frac{3}{16\pi} \frac{\lambda_p}{R}
    +\frac{9}{64\pi^3}  \frac{\lambda_p^3}{R^3} - \frac{9}{32\pi^2}
\frac{\lambda_p^2}{R^2} \coth\left(2\pi
\frac{R}{\lambda_p}\right) \right] \frac{R^4}{d^4} \right.\\
& + {\cal O} \left((R/d)^5\right)\Bigg\}
  \end{split}
\end{equation}
which applies for $d\gg R$, $\lambda_p$, $\lambda_T=\hbar c /k_B T$.
Notice that this energy is not universal in the sense that the
leading term depends on the plasma wave length. For $\lambda_p \ll R$,
the amplitude of the leading term becomes $-3/8$, in agreement with
the high-temperature result for perfect reflectors
\cite{Canaguier-Durand:2010fk}.  For $\lambda_p \gg R$, the amplitude
of the leading term approaches $-1/4$ which is identical to the result
for the Drude model (see Eq.~(\ref{eq:energy_high_T_plasma}) below).
The behavior in these two limits is consistent with the corresponding
limits of the zero-temperature result of Eq.~(\ref{eq:energy-plasma}).

\subsection{Drude model}

For this model, the T-matrix of the plane for $\kappa\to 0$ at {\it
  fixed} $k_\|$ behaves differently from
Eq.~\eqref{eq:T-plate_Drude_large_d}.  While $T_p^{\textsc{E}}\to 1$,
the magnetic part vanishes, $T_p^{\textsc{M}}\to 0$.
Eq.~\eqref{eq:U-matrix} shows that to leading order for small
$\kappa$, the matrix elements $U^{EE}_{lml'm}$ are given by
Eq.~\eqref{eq:U_small_kappa}.  In fact, we do not need to find the
other matrix elements of $U$: the elements coupling unlike
polarizations are reduced by a factor $\kappa$, and the elements
$U^{MM}_{lml'm}$ are multiplied by $T^M_{s,lm}$ of the sphere, which
scales as $\kappa^{2l+2}$ for small values of $\kappa$, and are thus
smaller by a factor $\kappa$ also. The (universal) elements of
$T^E_{s,lm}$ for small $\kappa$ are given by
Eq.~\eqref{eq:T_E_small_kappa}.  This shows that only the E
polarization contributes to the energy at high temperatures and from
Eqs.~\eqref{eq:U_small_kappa} and \eqref{eq:T_E_small_kappa} follows
the explicit result for the elements of the rescaled matrix $\tilde
Y^{EE}$,
\begin{widetext}
\begin{equation}
  \label{eq:Y_EE_small_k}
  \lim_{u\to 0}\tilde Y^{EE}_{lml'm}(u) = \frac{l+1}{l} \frac{1}{2^{l+l'+1}}
\sqrt{\frac{l(2l+1)l'(2l'+1)}{(l+1)(l'+1)}}\frac{(2l'-1)!!(l+l')!}
{(2l+1)!!} \frac{(R/d)^{2l+1}}{\sqrt{(l+m)!(l-m)!(l'+m)!(l'-m)!}} \, .
\end{equation}
\end{widetext}
In the high temperature limit the energy is then given by
\begin{equation}
  \label{eq:energy_high_T_drude}
  E = \frac{k_B T}{2}   \log \det 
[1-\tilde Y^{EE}(u\to 0) ] \, .
\end{equation}
Notice that this result is universal at {\it all} separations since
the matrix $\tilde Y$ depends only on $R/d$. The absence of magnetic
contributions is in agreement with the high temperature interaction
between two parallel plates that are described by the Drude model.  A
truncation of the matrix $\tilde Y$ at $l=2$ and expansion of
Eq.~\eqref{eq:energy_high_T_drude} for small $R/d$ yields
the large distance result
\begin{equation}
  \label{eq:energy_high_T_drude_large_d}
  E = - k_B T \left[ \frac{1}{4} \left(\frac{R}{d}\right)^3
    \!\!+\frac{1}{4} \left(\frac{R}{d}\right)^5 \!\!+ \frac{3}{128}
    \left(\frac{R}{d}\right)^6 \!\!+ {\cal O}\left(\frac{1}{d^7}\right)\right] 
\end{equation}
which applies when $d \gg R$, $\lambda_T$.

\section{Numerics}
\label{sec:numerics}

In this section, we evaluate the Casimir energy based on
Eq.~\eqref{eq:energy_general_T0} for zero temperature and
Eq.~\eqref{eq:energy_general} for finite temperatures.  Our results
are obtained by numerical computation of the determinant, the
integral over $\kappa$ (or sum over $n$) and the integral over $z$ of
Eq.~\eqref{eq:U-matrix}. The matrix $Y$ is truncated at a finite
partial wave order, $\ell_{max}$. We chose $\ell_{max}$ such that the
result for the energy changes by less than a factor
of $1.0001$ upon increasing $\ell_{max}$ by $10$. The required value
of $\ell_{max}$ depends on the separation between the plate and
sphere.  As the separation decreases, $\ell_{max}$ increases.  For
example for $R/d < 0.75$, we used $\ell_{max}=24$, whereas for $R/d
=0.8$ and $0.85$, one needs $\ell_{max}=34$ and for $R/d=0.9$ the
value $\ell_{max}=54$.

The numerical computation of the determinant, the integrals and sum
poses no principle problem. However, it is important to consider the
determinant of Eq.~\eqref{eq:energy_general_T0} or
Eq.~\eqref{eq:energy_general} carefully for $\kappa \to 0$.  In
Sect.~\ref{sec:kappa=0} we have already seen that the matrix elements
$Y_{lml'm}$ for small $\kappa$ scale as $\kappa^{l-l'}$ or
$\kappa^{l-l'+1}$. This shows that for small $\kappa$ the matrix
elements with $l\gg l'$ become extremely small whereas those with $l
\ll l'$ increase rapidly.  For large values of $l_{max}$ this behavior
makes the computation of the determinant at $\kappa=0$ numerically
ill-conditioned.  However, the analytical results presented in
Sect.~\ref{sec:kappa=0} allow us to calculate the $n=0$ term in
Eq.~\eqref{eq:energy_general} or the integrand of
Eq.~\eqref{eq:energy_general_T0} at $\kappa=0$ for $\ell_{max}=100$
and even larger.  In fact, as the value of $R/d$ is increased beyond
$0.9$, larger values $\ell_{max}>60$ must be used in order to
accurately calculate the energy.  For sufficiently high temperatures,
the second Matsubara wave vector $\kappa_1=2\pi k_B T/\hbar c$ in
Eq.~\eqref{eq:energy_general} becomes sufficiently large and hence
poses no numerical problem for the computation of the energy. For
example, for $T=300^\circ K$ the Casimir energy can be calculated for
$R/d=0.95$ with $\ell_{max}=72$.  As $\ell_{max}$ increases, the
interval in the vicinity of $\kappa=0$ in which the integrand cannot
be obtained with sufficient precision numerically increases too. Due
to this behavior, we restrict the calculation of the Casimir energy at
$T=0$ to $R/d\leq 0.9$.

\subsection{Casimir interaction at $T=0$}

In this section, we calculate the Casimir energies for the usual Drude
and plasma model given in Eqs.~\eqref{eq:eps_plate} and
\eqref{eq:eps_Drude}, respectively, for parameters of gold as
given below.  There are three dimensionless parameters which we choose
as $d/R$, $\lambda_p/R$ and $\lambda_p/\lambda_\sigma$. The first two
parameters can be controlled for a given material by changing the
separation $d$ and the radius $R$ of the sphere. In order to avoid
strong finite size effects in the electronic response, we assume that
$\lambda_\sigma$, $\lambda_p< R$.

\begin{figure}[ptb]
\includegraphics[width=1\columnwidth]{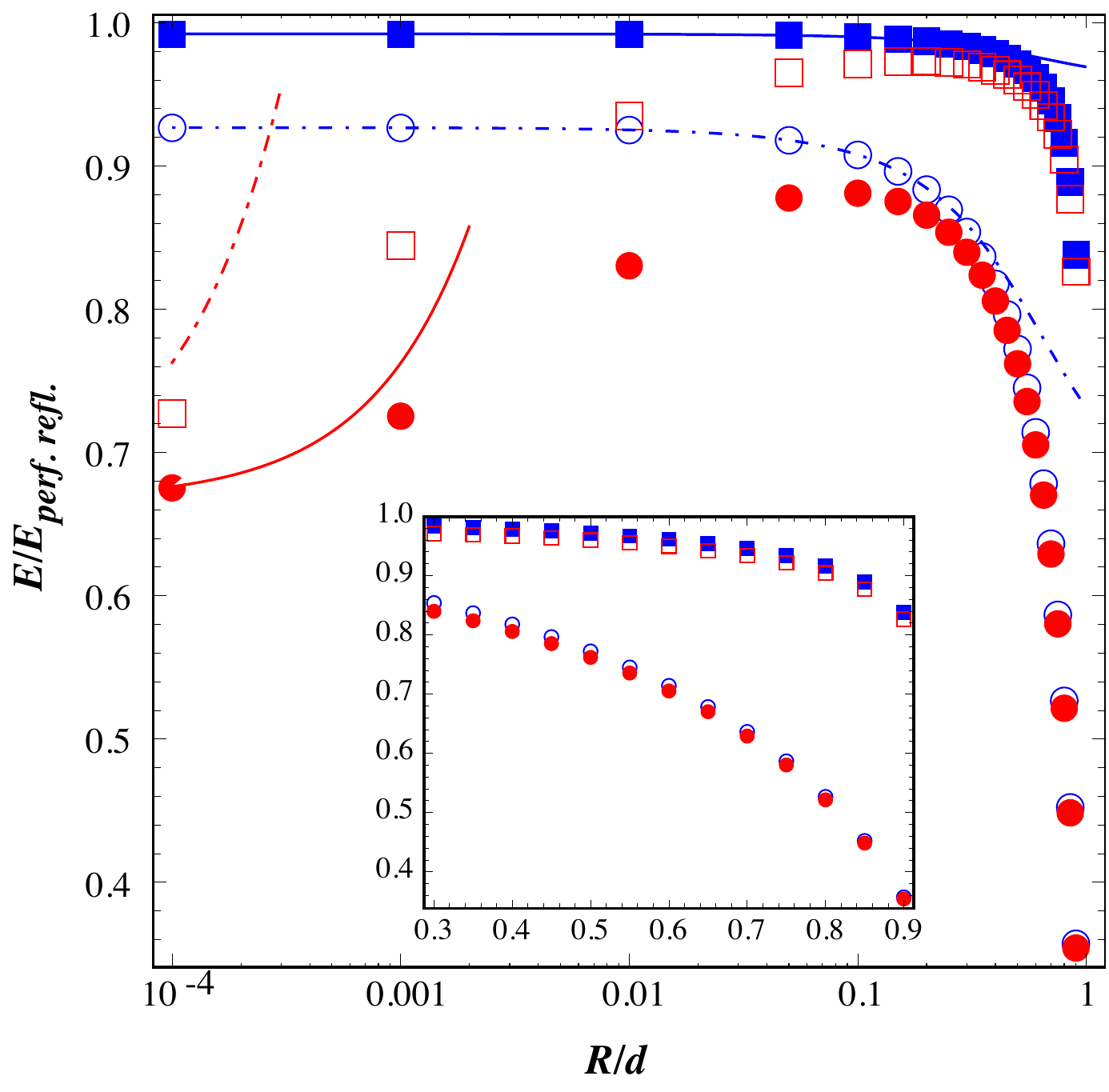}
\caption{$E/E_{\rm perf. refl.}$ against $R/d$ for the plasma model
  for $\lambda_p/R=0.5$, $0.05$ (open circles and filled squares
  respetively), and the Drude model for the same values of
  $\lambda_p/R$ (filled circles and open squares, respectively) and
  $\lambda_p/\lambda_\sigma= 27.4$. The solid and dashed lines
  represent the asymptotic results of Eqs.~\eqref{eq:energy-plasma},
  \eqref{eq:E_Drude_both} for the plasma and the Drude model,
  respectively. Inset: Magnification of the short distance range.}
\label{fig:comp}
\end{figure}

In Sections \ref{sec:plasma_exact} and \ref{sec:drude_exact}, using
$\ell_{max}=2$ partial waves, we obtained an asymptotic expansion of
the Casimir energy for both plasma and Drude model at large
separations, see Eqs.~\eqref{eq:energy-plasma} and
\eqref{eq:E_Drude_both}.  In Fig.~\ref{fig:comp}, we compare the
analytical results to the exact numerical results that were obtained
as described before.  The graph shows the exact energies for the Drude
and the plasma model normalized to the exact energies for perfect
reflecting surfaces, taken from Ref.\cite{Emig:2008ee}.  For the
plasma model we used $\lambda_p/R=0.05$ and $0.5$, respectively, and
for the Drude model the same two values for $\lambda_p/R$ and we set
$\lambda_p/\lambda_{\sigma}=27.4$.  The figure illustrates the
material dependence of the Casimir energies.  For large separations,
the ratios for the plasma model approach values slightly smaller than one,
which is consistent with the $\lambda_p/R$-dependent asymptotic form
predicted by Eq.~(\ref{eq:energy-plasma}).  For the case of the Drude
model, the ratio tends to the universal number $2/3$ at large
separations, as predicted by Eqs.~(\ref{eq:E_perf_large_d}),
(\ref{eq:E_Drude_both}).  In the case of the plasma model, the asymptotic
result describes the energy up to $R/d\approx 0.4$ nicely. For the
Drude model, however, the agreement between the analytical and
numerical findings is limited to extremely small $R/d\lesssim
10^{-4}$. This example clearly indicates distinct correlations between
material and geometry. Our result shows that for the Drude model a
larger number of partial waves than for the plasma model is necessary
to accurately calculate the Casimir energy.

\begin{figure}[ptb]
\includegraphics[width=1\columnwidth]{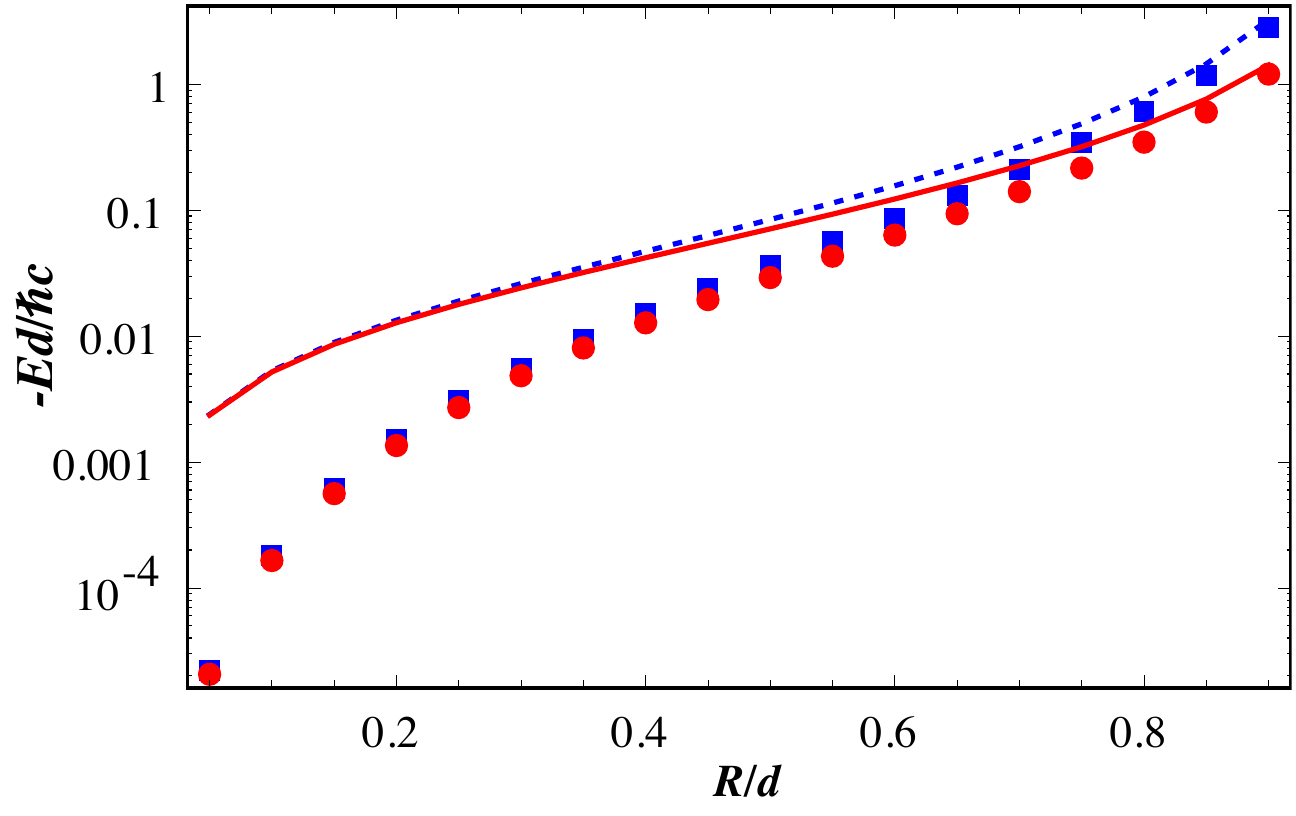}
\caption{Numerical result for $-E d/\hbar c$ against $R/d$ for the
  plasma model at $\lambda_p/R=0.05$ (squares) and at
  $\lambda_p/R=0.5$ (circles).  The lines represent the PFA energy at
  $\lambda_p/R=0.05$ (dashed) and at $\lambda_p/R=0.5$ (solid). }
\label{fig:pfa}
\end{figure}

\begin{figure}[ptb]
\includegraphics[width=0.9\columnwidth]{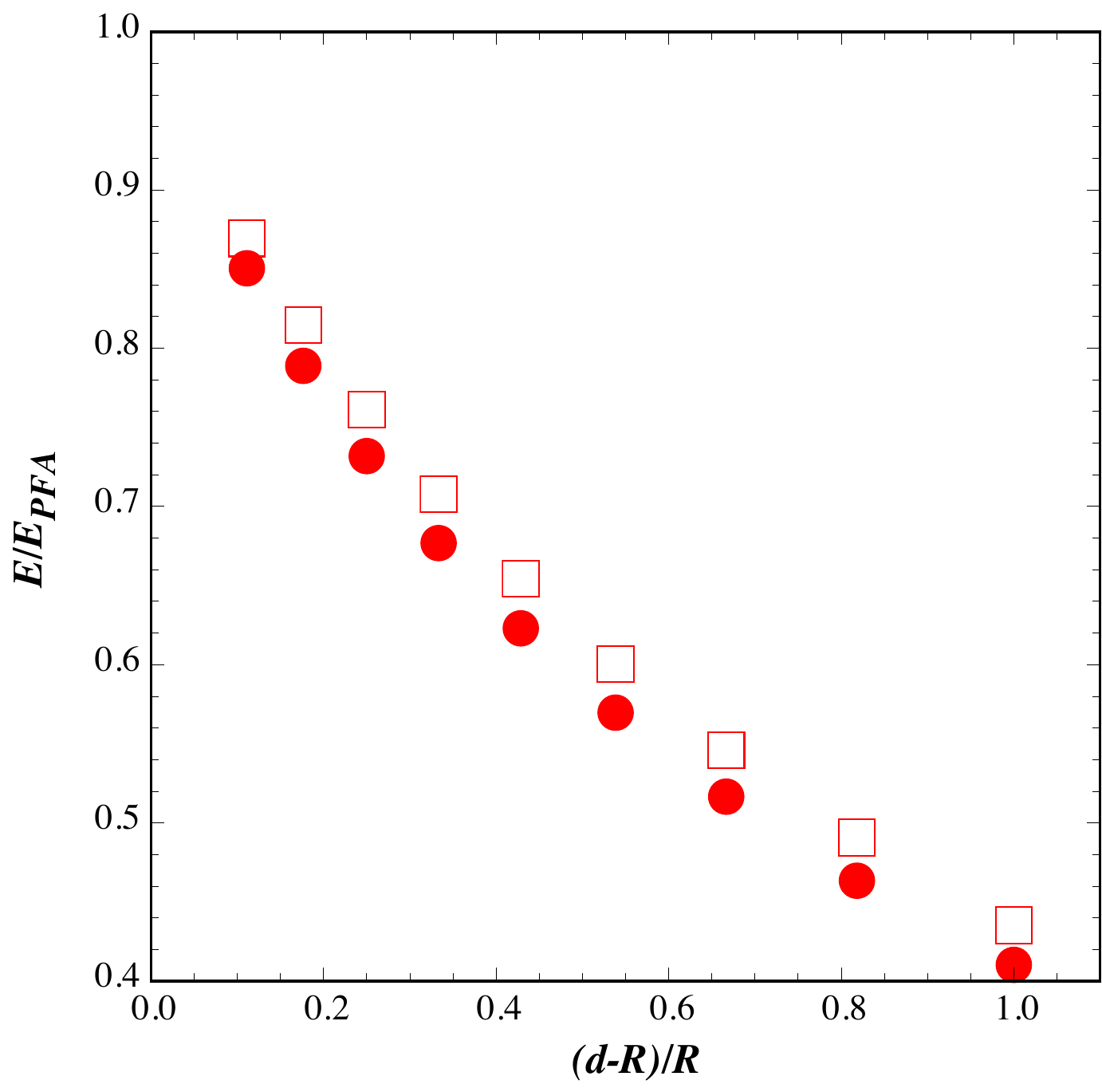}
\caption{Ratio of the numerical results for the Casimir energy shown
  in Fig.~\ref{fig:pfa} and the PFA energy based on the Lifshitz
  theory for the plasma model with $\lambda_p/R=0.05$ (squares) and
  $\lambda_p/R=0.5$ (circles). The ratio is shown as a function of the
  surface-to-surface distance $d-R$.}
\label{fig:pfa_corr}
\end{figure}

We also compare the exact numerical results with the Casimir energy
obtained by the PFA for both the plasma and the Drude model.  The PFA
energy is obtained by integrating the PFA force $F=2\pi R E_{\rm
  plates}(d-R)$ with respect to $d$, where $E_{\rm plates}(d-R)$ is
the energy of two parallel plates at distance $d-R$ as given by the
Lifshitz formula \cite{Lifshitz56} with the corresponding dielectric
function.  Fig.~\ref{fig:pfa} shows the exact Casimir energy
calculated numerically for the plasma model with plasma wavelength
$\lambda_p=0.05$ and $\lambda_p=0.5$, respectively.  The figure shows
also the PFA energy for the same values of $\lambda_p$. As expected,
the discrepancy between the exact and PFA energy decreases as $R/d$
increases and is expected to vanish for $d\to R$. This is clearly
visible from Fig.~\ref{fig:pfa_corr} which shows the relative
corrections to the PFA energy at short separations. Interestingly, the
dependence of the corrections on $\lambda_p$ is not fully described by
the Lifshitz theory since the data for different $\lambda_p$ do not
collapse onto a single curve. This demonstrates correlations between
geometry and material properties that are not described by the PFA.
For example, for $\lambda_p/R=0.5$ and $\lambda_p/R=0.05$ we find at
the shortest studied separation of $d-R=0.11 R$ the exact energy to be
$85\%$ and $87\%$ of the PFA energy, respectively. For perfect
reflectors the reduction was found to be $\approx
87\%$ at the same distance \cite{Emig:2008ee}.  We find similar results using the Drude
model. The energies associated with the Drude model are not shown here
since they collapse onto the data for the plasma model at short
separations.

\subsection{Casimir interaction at $T\neq 0$}

The Casimir free energy at finite temperatures $T$ depends on the
thermal wavelength $\lambda_T=\hbar c/k_BT$.  This additional length
scale introduces an additional dimensionless parameter $\lambda_T/R$.
To investigate the influence of temperature, we calculated the Casimir
free energy at two different values of this parameter. We have chosen
the values $\lambda_T/R=1.52$, $5.94$ since they correspond, e.g., to
a sphere of radius $R=5 \mu m$, which is small but still relevant to
experiments. The temperature is chosen as $T=300^\circ K$ and $T=77^\circ
K$, yielding $\lambda_T=7.6\mu$m and $\lambda_T=29.7\mu$m,
respectively.  These two temperatures, corresponding to room
temperature and the boiling point of molecular nitrogen N$_2$
respectively, can readily be accessed in experiments. 

Below, we employ more detailed models for the material response to
calculate the Casimir energies at higher temperatures. More
specifically, we consider generalized plasma and Drude models, which
take into account the interband transitions of core electrons that
are described by a set of oscillators with nonzero resonant
frequencies.  The generalized plasma model has the dielectric
permittivity
\begin{equation}
  \label{eq:eps_gen_Pl}
  \epsilon_p(ic\kappa) = 1+
  \left(\frac{2\pi}{\lambda_p\kappa}\right)^2 +\epsilon_c(ic\kappa)
\end{equation}
and the generalized Drude is described by 
\begin{equation}
\label{eq:eps_gen_Drude}
  \epsilon_p(ic\kappa)=1+\frac{(2\pi)^2}{(\lambda_p\kappa)^2+\pi c \kappa/\sigma_p} +\epsilon_c(ic\kappa)
\end{equation}
with
\begin{equation}
\label{eq:A_omega}
 \epsilon_c(ic\kappa)=\sum_{j=1}^{K}\frac{f_j}{\omega_j^2+(c\kappa)^2+g_jc\kappa}
\end{equation}
Here $K$ is the number of oscillators, $f_j$ are the oscillator
strengths, $g_j$ are the relaxation frequency and $\omega_j \neq 0$ are the
resonant frequencies of the core electrons. Typical parameters for
gold are given by $\omega_p=9$eV for the plasma frequency and
$\gamma=35$meV for the relaxation rate \cite{Canaguier09}. These
parameters correspond to the length scales $\lambda_p=2\pi c/\omega_p
= 137$nm and $\lambda_\gamma=2\pi c/\gamma=35.4\mu$m. For these
parameters, the dc conductivity $\sigma=\omega_p^2/(4\pi \gamma)$ is
$184.2$eV, corresponding to the length scale $\lambda_\sigma=2\pi
c/\sigma=6.7$nm.  Note that electron scattering is not described by
the usual plasma model. However, as can be seen from
Eq.~\eqref{eq:eps_gen_Pl}, dissipation is included in the generalized
plasma model due to the interband transition of core electrons. To
calculate the Casimir energy, we use the oscillator parameters of gold
which are presented in Table \ref{tbl:generalized}.  These parameters
have been calculated \cite{Decca07} based on the 6-oscillator model
fitted to the tabulated optical data given in Ref.~\cite{Palik:85}. 
\begin{table}[h]
\begin{center}
\begin{tabular}{c|c|c|c|c|c}
$j$   &$ \omega_j\,[eV]$      & $g_j\,[eV] $		&$ f_j\, [eV^2]$ \\
\hline \hline
$1$ & $3.05$ & $0.75$ 			& $7.091$   \\
$2$ & $4.15$ 	& $1.85$ 			& $41.46$\\
$3$ & $5.4$ 	& 1.0 & $2.700$ \\
$4$ & $8.5$ & $7.0$ 			& $154.7$  \\
$5$ & $13.5$ & $6.0$ 			& $44.55$ \\
$6$ & $21.5$& $9.0$ 			& $309.6$  \\
\hline
\end{tabular}
\end{center}
\caption{Oscillator parameters for gold. Calculated in
  Ref. \cite{Decca07} by  fitting 6 oscillators to 
  tabulated optical data \cite{Palik:85}.}
\label{tbl:generalized}
\end{table}

We first calculated the $n=0$ term of the sum of
Eq.~\eqref{eq:energy_general} analytically, based on the expressions
given in Section \ref{sec:kappa=0} and then calculate the other terms
numerically as explained previously.  This allows us to calculate the
Casimir free energy for very short separations. As explained above,
large values of $\ell_{max}$ should be used at short separations but
in this limit the numerical evaluation of the determinant in
Eq.~\eqref{eq:energy_general} is cumbersome. This problem disappears
as the temperature in increased because then the second Matsubara wave
vector $\kappa_1$ becomes larger and thus no calculation for too small
values of $\kappa>0$ will be necessary. For the purpose of this paper,
we calculate the relevant energies for distances as short as
$R/d=0.95$ at the two different temperatures using the generalized
form of the Drude and the plasma model, see
Eqs.~\eqref{eq:eps_gen_Pl}, \eqref{eq:eps_gen_Drude}.

\begin{figure}[ptb]
\includegraphics[width=1\columnwidth]{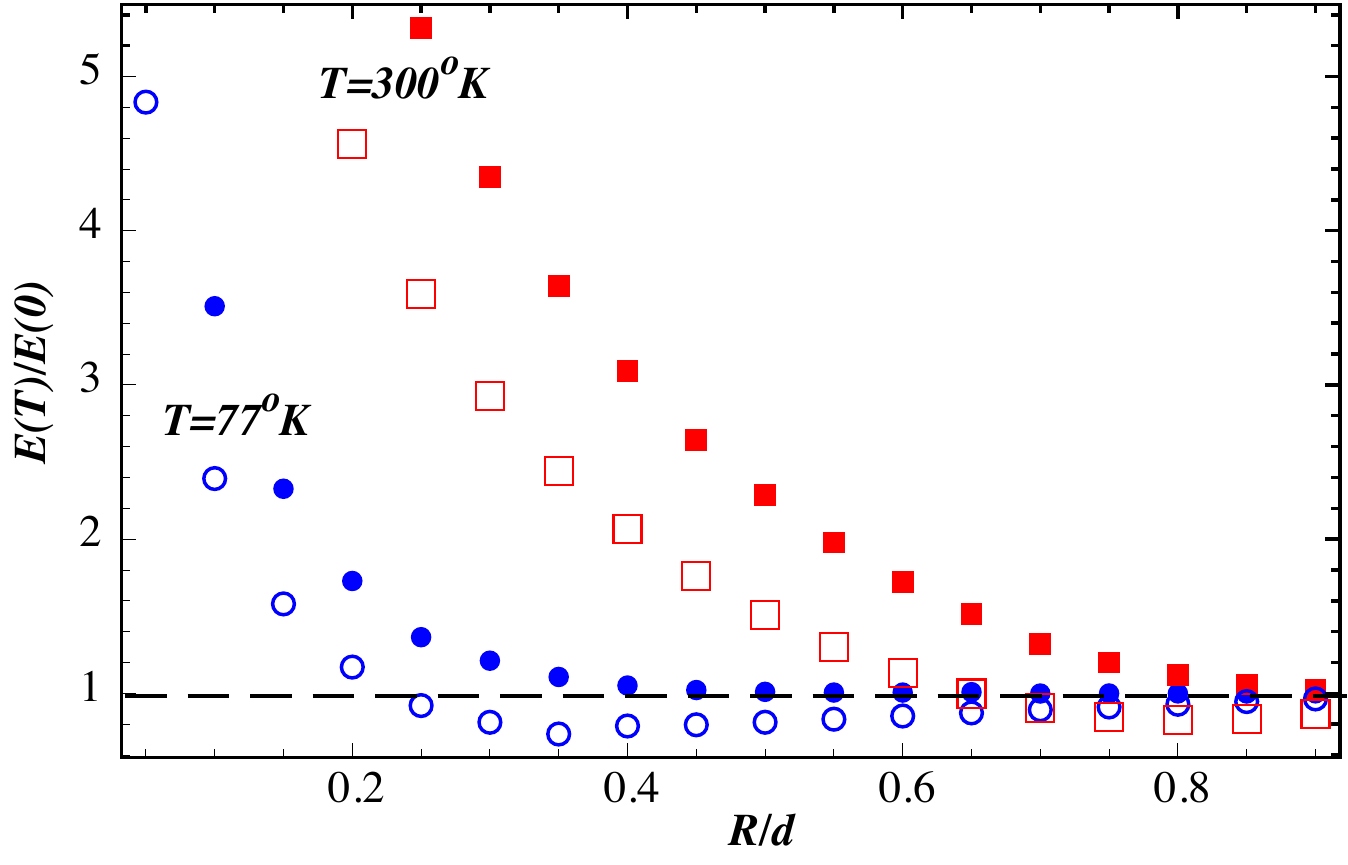}
\caption{Ratio of the Casimir free energy to the energy at $T=0$
  against $R/d$ for the generalized Drude model (open symbols) and the
  generalized plasma model (filled symbols) at $T=300^\circ K$
  (squares) and at $T=77^\circ K$ (circles) for a sphere of radius
  $R=5\mu$m.}
\label{fig:ratio}
\end{figure}

Fig.~\ref{fig:ratio} shows the ratio of the Casimir free energy to the
energy at $T=0$ for the generalized plasma model and for the
generalized Drude model at $T=300^\circ K$ and $T=77^\circ K$. While
the Casimir free energy at $T=300^\circ K$ is always larger than at
$T=77^\circ K$ for the plasma model, we find that for the Drude model
the Casimir free energies at $T=77^\circ K$ and $T=300^\circ K$ cross
each other around $R/d=0.7$.  For $R/d \gtrsim 0.7$, the Casimir free
energy corresponding to $T=77^\circ K$ becomes larger than the one for
$T=300^\circ K$. For the sphere plate geometry, indications of
negative entropy have recently been reported
\cite{Canaguier-Durand:2010fk}. The ratio shown in
Fig.~\ref{fig:ratio} can be expressed as $E(T)/E(0)=1-TS/E(0)$ where
$S$ is the entropy associated with the Casimir free energy
$E(T)$. Hence, a ratio $E(T)/E(0)<1$ implies a negative entropy since
$E(0)<0$.  Our results clearly show that for the Drude model the
entropy indeed becomes negative for sufficiently small separations.
However, for the plasma model our results of Fig.~\ref{fig:ratio}
indicate a positive entropy.

\begin{figure}[ptb]
\includegraphics[width=1\columnwidth]{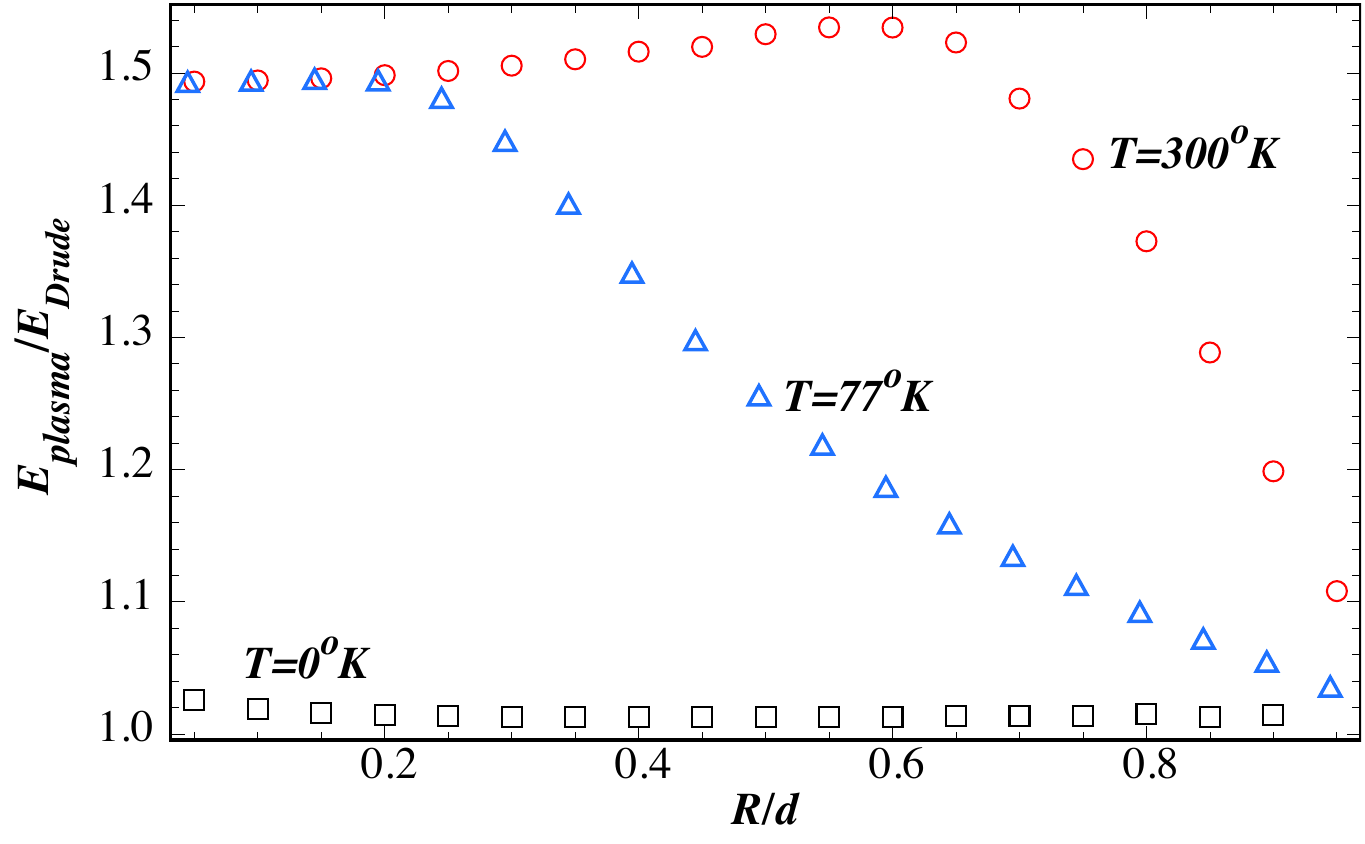}
\caption{The ratio of Casimir free energies for the generalized plasma
  and generalized Drude model at $T=300^\circ K$ (circles),
  $T=77^\circ K$ (triangles) and $T=0$ (squares).}
\label{fig:ratio_to_T0}
\end{figure}

Above we showed that at large separations the ratio of the Casimir
energy for the plasma and the Drude model varies between $3/2$ (for
small $\lambda_p/R$) and $1$ (for large $\lambda_p/R$) for zero and
finite temperatures.  At shorter separations the ratio is expected to
tend to one since at high frequencies the plasma and Drude model
become identical.  It is interesting to observe how the ratio tends to
one with decreasing separation as a function of
temperature. Fig.~\ref{fig:ratio_to_T0} shows the Casimir free energy
for the plasma model divided by that for the Drude model at
$T=300^\circ K$, $T=77^\circ K$ and $T=0^\circ K$. Since
$\lambda_p/R=0.0274$ is small compared to one, the ratio tends to
almost $3/2$ at large separations.  As shown in the figure, with
decreasing separation the ratio drops towards one very fast for
$T=0$. However, for $T=300^\circ K$ the ratio is larger than $3/2$ for
$R/d\lesssim 0.7$, goes through a maximum around $R/d=0.6$ and
finally starts dropping to one. The curve for $T=77^\circ K$ also
displays a slight maximum close to $R/d=0.15$.  A similar behavior
with an extrema has been observed also in
Ref.~\cite{Canaguier-Durand:2010fk} for a sufficiently large sphere at
$T=300^\circ K$. The maxima occur at a distance that approximately
corresponds to the thermal wavelength $\lambda_T$ with
$R/\lambda_T=0.66$ and $R/\lambda_T=0.17$ for $T=300^\circ K$ and
$T=77^\circ K$, respectively. Since thermal photons of wavelength
$\lambda_T$ mostly contribute to the energy at a separation $d\approx
\lambda_T$, the position of the maxima suggests that thermal effects
less strongly enhance the Drude energy than the plasma energy,
presumably due to dissipation.

\section{Conclusion}

We have shown in detail how the scattering approach for Casimir
interactions can be applied to study correlations between effects due
to geometry, material properties and finite temperature. The
experimentally most relevant geometry of a sphere and plate reveals
interesting properties of the Casimir interaction that are absent for
parallel plates and hence in the proximity force approximation. These
findings demonstrate an interplay between material properties and the
finite size of the sphere. Our main results are as follows. At large
separations we observe both at zero and finite temperatures for the
amplitude of the leading term of the energy different results for
perfect reflectors, Drude and plasma metals. The plasma model yields a
non-universal amplitude that depends on the ratio of plasma wavelength
to sphere radius. For the perfect reflector and Drude model the
amplitudes are universal but for the latter it is reduced by a factor of
$2/3$. This result is distinct from the interaction of two parallel
plates, which at zero temperature is asymptotically identical for the
three material descriptions. The identification of the plasma
wavelength with the penetration depth of a London superconductor
explains why the plasma model yields the asymptotic interaction for
perfect reflectors and a Drude metals as limiting cases.

Our numerical computations of the energy at smaller separations
demonstrate further important differences between the plasma and Drude
model and generalizations thereof.  We observe full agreement of the
numerical results with the asymptotic expansion at large separations
that, however, is limited for the Drude model to extremely large
distances. Hence, we conclude that for the Drude model higher order
multipoles are more important than for the plasma model. At small
separations the observed dependence of the difference between the
exact and the PFA energies on the plasma wavelength demonstrates that
geometry and material effects are correlated. Our results at finite
temperatures show that the Casimir energy for Drude metals changes
non-monotonically with temperature, leading to a larger energy at
$T=77^\circ$K than at $T=300^\circ$K at sufficiently small
separations. We observe a negative entropy associated with the Casimir
free energy for the Drude model over a range of distances. This range
increases when the temperature is decreased.  Both non-monotonic
temperature dependence and negative entropy are not observed for the
plasma model in the range of studied parameters. At finite
temperatures, we find that the Casimir free energy for plasma metals
is approximately $3/2$ times the energy for Drude metals for
separations $d\gtrsim \lambda_T$.

\acknowledgements The results for large separations at $T=0$ and their
interpretation in terms of a London superconductor have been presented
before at the QFEXT09 conference at the University of Oklahoma.
During the completion of this work we became aware of a related work
that deals with the sphere-plate interaction at $T=300^\circ
K$\cite{Canaguier-Durand:2010fk}. We thank G.~Bimonte, M.~Kardar for
useful conversations regarding this work. We are grateful to
S.~J.~Rahi for his extensive help on various aspects of this work.
This work was supported by the National Science Foundation (NSF)
through grants DMR-06-45668 (RZ), Defense Advanced Research Projects
Agency (DARPA) contract No. S-000354 (RZ, TE and UM), and by the
Deutsche Forschungsgemeinschaft (DFG) through grant EM70/3 (TE).

\bibliographystyle{apsrev}
\bibliography{casimir}

\begin{thebibliography}{23}
\expandafter\ifx\csname natexlab\endcsname\relax\def\natexlab#1{#1}\fi
\expandafter\ifx\csname bibnamefont\endcsname\relax
  \def\bibnamefont#1{#1}\fi
\expandafter\ifx\csname bibfnamefont\endcsname\relax
  \def\bibfnamefont#1{#1}\fi
\expandafter\ifx\csname citenamefont\endcsname\relax
  \def\citenamefont#1{#1}\fi
\expandafter\ifx\csname url\endcsname\relax
  \def\url#1{\texttt{#1}}\fi
\expandafter\ifx\csname urlprefix\endcsname\relax\def\urlprefix{URL }\fi
\providecommand{\bibinfo}[2]{#2}
\providecommand{\eprint}[2][]{\url{#2}}

\bibitem[{\citenamefont{Lamoreaux}(1997)}]{Lamoreaux97}
\bibinfo{author}{\bibfnamefont{S.~K.} \bibnamefont{Lamoreaux}},
  \bibinfo{journal}{Phys. Rev. Lett.} \textbf{\bibinfo{volume}{78}},
  \bibinfo{pages}{5 } (\bibinfo{year}{1997}).

\bibitem[{\citenamefont{Mohideen and Roy}(1998)}]{Mohideen98}
\bibinfo{author}{\bibfnamefont{U.}~\bibnamefont{Mohideen}} \bibnamefont{and}
  \bibinfo{author}{\bibfnamefont{A.}~\bibnamefont{Roy}},
  \bibinfo{journal}{Phys. Rev. Lett.} \textbf{\bibinfo{volume}{81}},
  \bibinfo{pages}{4549 } (\bibinfo{year}{1998}).

\bibitem[{\citenamefont{Harris et~al.}(2000)\citenamefont{Harris, Chen, and
  Mohideen}}]{Harris:2000a}
\bibinfo{author}{\bibfnamefont{B.~W.} \bibnamefont{Harris}},
  \bibinfo{author}{\bibfnamefont{F.}~\bibnamefont{Chen}}, \bibnamefont{and}
  \bibinfo{author}{\bibfnamefont{U.}~\bibnamefont{Mohideen}},
  \bibinfo{journal}{Phys. Rev. A} \textbf{\bibinfo{volume}{62}},
  \bibinfo{pages}{052109} (\bibinfo{year}{2000}).

\bibitem[{\citenamefont{Decca et~al.}(2005)\citenamefont{Decca, Lopez,
  Fischbach, Klimchitskaya, Krause, and Mostepanenko}}]{Decca:2005b}
\bibinfo{author}{\bibfnamefont{R.~S.} \bibnamefont{Decca}},
  \bibinfo{author}{\bibfnamefont{D.}~\bibnamefont{Lopez}},
  \bibinfo{author}{\bibfnamefont{E.}~\bibnamefont{Fischbach}},
  \bibinfo{author}{\bibfnamefont{G.~L.} \bibnamefont{Klimchitskaya}},
  \bibinfo{author}{\bibfnamefont{D.~E.} \bibnamefont{Krause}},
  \bibnamefont{and} \bibinfo{author}{\bibfnamefont{V.~M.}
  \bibnamefont{Mostepanenko}}, \bibinfo{journal}{Ann. Phys. (N.Y.)}
  \textbf{\bibinfo{volume}{318}}, \bibinfo{pages}{37} (\bibinfo{year}{2005}).

\bibitem[{\citenamefont{Decca et~al.}(2007)\citenamefont{Decca, L\'{o}pez,
  Fischbach, Klimchitskaya, Krause, and Mostepanenko}}]{Decca07}
\bibinfo{author}{\bibfnamefont{R.~S.} \bibnamefont{Decca}},
  \bibinfo{author}{\bibfnamefont{D.}~\bibnamefont{L\'{o}pez}},
  \bibinfo{author}{\bibfnamefont{E.}~\bibnamefont{Fischbach}},
  \bibinfo{author}{\bibfnamefont{G.~L.} \bibnamefont{Klimchitskaya}},
  \bibinfo{author}{\bibfnamefont{D.~E.} \bibnamefont{Krause}},
  \bibnamefont{and} \bibinfo{author}{\bibfnamefont{V.~M.}
  \bibnamefont{Mostepanenko}}, \bibinfo{journal}{Phys. Rev. D}
  \textbf{\bibinfo{volume}{75}}, \bibinfo{eid}{077101} (\bibinfo{year}{2007}).

\bibitem[{\citenamefont{Klimchitskaya et~al.}(2009)\citenamefont{Klimchitskaya,
  Mohideen, and Mostepanenko}}]{Klimchitskaya:2009vn}
\bibinfo{author}{\bibfnamefont{G.~L.} \bibnamefont{Klimchitskaya}},
  \bibinfo{author}{\bibfnamefont{U.}~\bibnamefont{Mohideen}}, \bibnamefont{and}
  \bibinfo{author}{\bibfnamefont{V.~M.} \bibnamefont{Mostepanenko}},
  \bibinfo{journal}{Rev. Mod. Phys.} \textbf{\bibinfo{volume}{81}},
  \bibinfo{pages}{1827} (\bibinfo{year}{2009}).

\bibitem[{\citenamefont{Casimir}(1948)}]{Casimir48-1}
\bibinfo{author}{\bibfnamefont{H.~B.~G.} \bibnamefont{Casimir}},
  \bibinfo{journal}{Proc. K. Ned. Akad. Wet.} \textbf{\bibinfo{volume}{51}},
  \bibinfo{pages}{793} (\bibinfo{year}{1948}).

\bibitem[{\citenamefont{Dzyaloshinskii
  et~al.}(1961)\citenamefont{Dzyaloshinskii, Lifshitz, and
  Pitaevskii}}]{Dzyaloshinskii61}
\bibinfo{author}{\bibfnamefont{I.~E.} \bibnamefont{Dzyaloshinskii}},
  \bibinfo{author}{\bibfnamefont{E.~M.} \bibnamefont{Lifshitz}},
  \bibnamefont{and} \bibinfo{author}{\bibfnamefont{L.~P.}
  \bibnamefont{Pitaevskii}}, \bibinfo{journal}{Adv. Phys.}
  \textbf{\bibinfo{volume}{10}}, \bibinfo{pages}{165} (\bibinfo{year}{1961}).

\bibitem[{\citenamefont{Krause et~al.}(2007)\citenamefont{Krause, Decca,
  L\'{o}pez, and Fischbach}}]{Krause:2007a}
\bibinfo{author}{\bibfnamefont{D.~E.} \bibnamefont{Krause}},
  \bibinfo{author}{\bibfnamefont{R.~S.} \bibnamefont{Decca}},
  \bibinfo{author}{\bibfnamefont{D.}~\bibnamefont{L\'{o}pez}},
  \bibnamefont{and}
  \bibinfo{author}{\bibfnamefont{E.}~\bibnamefont{Fischbach}},
  \bibinfo{journal}{Phys. Rev. Lett.} \textbf{\bibinfo{volume}{98}},
  \bibinfo{eid}{050403} (\bibinfo{year}{2007}).

\bibitem[{\citenamefont{Emig et~al.}(2009)\citenamefont{Emig, Graham, Jaffe,
  and Kardar}}]{Emig:2009kx}
\bibinfo{author}{\bibfnamefont{T.}~\bibnamefont{Emig}},
  \bibinfo{author}{\bibfnamefont{N.}~\bibnamefont{Graham}},
  \bibinfo{author}{\bibfnamefont{R.~L.} \bibnamefont{Jaffe}}, \bibnamefont{and}
  \bibinfo{author}{\bibfnamefont{M.}~\bibnamefont{Kardar}},
  \bibinfo{journal}{Phys. Rev. A} \textbf{\bibinfo{volume}{79}},
  \bibinfo{pages}{054901} (\bibinfo{year}{2009}).

\bibitem[{\citenamefont{Rahi et~al.}(2009)\citenamefont{Rahi, Emig, Graham,
  Jaffe, and Kardar}}]{rahi09}
\bibinfo{author}{\bibfnamefont{S.~J.} \bibnamefont{Rahi}},
  \bibinfo{author}{\bibfnamefont{T.}~\bibnamefont{Emig}},
  \bibinfo{author}{\bibfnamefont{N.}~\bibnamefont{Graham}},
  \bibinfo{author}{\bibfnamefont{R.~L.} \bibnamefont{Jaffe}}, \bibnamefont{and}
  \bibinfo{author}{\bibfnamefont{M.}~\bibnamefont{Kardar}},
  \bibinfo{journal}{Phys. Rev. D} \textbf{\bibinfo{volume}{80}},
  \bibinfo{eid}{085021} (\bibinfo{year}{2009}).

\bibitem[{\citenamefont{Bordag et~al.}(2009)\citenamefont{Bordag,
  Klimchitskaya, Mohideen, and Mostepanenko}}]{Bordag:2009uq}
\bibinfo{author}{\bibfnamefont{M.}~\bibnamefont{Bordag}},
  \bibinfo{author}{\bibfnamefont{G.~L.} \bibnamefont{Klimchitskaya}},
  \bibinfo{author}{\bibfnamefont{U.}~\bibnamefont{Mohideen}}, \bibnamefont{and}
  \bibinfo{author}{\bibfnamefont{V.~M.} \bibnamefont{Mostepanenko}},
  \emph{\bibinfo{title}{Advances in the Casimir effect}}
  (\bibinfo{publisher}{Oxford}, \bibinfo{year}{2009}).

\bibitem[{\citenamefont{Chen et~al.}(2007)\citenamefont{Chen, Klimchitskaya,
  Mostepanenko, and Mohideen}}]{Chen07}
\bibinfo{author}{\bibfnamefont{F.}~\bibnamefont{Chen}},
  \bibinfo{author}{\bibfnamefont{G.~L.} \bibnamefont{Klimchitskaya}},
  \bibinfo{author}{\bibfnamefont{V.~M.} \bibnamefont{Mostepanenko}},
  \bibnamefont{and} \bibinfo{author}{\bibfnamefont{U.}~\bibnamefont{Mohideen}},
  \bibinfo{journal}{Phys. Rev. B} \textbf{\bibinfo{volume}{76}},
  \bibinfo{eid}{035338} (\bibinfo{year}{2007}).

\bibitem[{\citenamefont{Parsegian}(2005)}]{Parsegian05}
\bibinfo{author}{\bibfnamefont{V.~A.} \bibnamefont{Parsegian}},
  \emph{\bibinfo{title}{van der {Waals Forces}}} (\bibinfo{publisher}{Cambridge
  University Press}, \bibinfo{address}{Cambridge}, \bibinfo{year}{2005}).

\bibitem[{\citenamefont{Emig}(2008)}]{Emig:2008ee}
\bibinfo{author}{\bibfnamefont{T.}~\bibnamefont{Emig}},
  \bibinfo{journal}{Journal of Statistical Mechanics: Theory and Experiment}
  \textbf{\bibinfo{volume}{4}}, \bibinfo{pages}{P04007} (\bibinfo{year}{2008}).

\bibitem[{\citenamefont{Canaguier-Durand
  et~al.}(2009)\citenamefont{Canaguier-Durand, Maia~Neto, Cavero-Pelaez,
  Lambrecht, and Reynaud}}]{Canaguier09}
\bibinfo{author}{\bibfnamefont{A.}~\bibnamefont{Canaguier-Durand}},
  \bibinfo{author}{\bibfnamefont{P.~A.} \bibnamefont{Maia~Neto}},
  \bibinfo{author}{\bibfnamefont{I.}~\bibnamefont{Cavero-Pelaez}},
  \bibinfo{author}{\bibfnamefont{A.}~\bibnamefont{Lambrecht}},
  \bibnamefont{and} \bibinfo{author}{\bibfnamefont{S.}~\bibnamefont{Reynaud}},
  \bibinfo{journal}{Phys. Rev. Lett.} \textbf{\bibinfo{volume}{102}},
  \bibinfo{pages}{230404} (\bibinfo{year}{2009}).

\bibitem[{\citenamefont{Canaguier-Durand
  et~al.}(2010)\citenamefont{Canaguier-Durand, Neto, Lambrecht, and
  Reynaud}}]{Canaguier-Durand:2010fk}
\bibinfo{author}{\bibfnamefont{A.}~\bibnamefont{Canaguier-Durand}},
  \bibinfo{author}{\bibfnamefont{P.~A.~M.} \bibnamefont{Neto}},
  \bibinfo{author}{\bibfnamefont{A.}~\bibnamefont{Lambrecht}},
  \bibnamefont{and} \bibinfo{author}{\bibfnamefont{S.}~\bibnamefont{Reynaud}},
  \bibinfo{journal}{Phys. Rev. Lett.} \textbf{\bibinfo{volume}{104}},
  \bibinfo{pages}{040403} (\bibinfo{year}{2010}).

\bibitem[{\citenamefont{Emig et~al.}(2006)\citenamefont{Emig, Jaffe, Kardar,
  and Scardicchio}}]{Emig06}
\bibinfo{author}{\bibfnamefont{T.}~\bibnamefont{Emig}},
  \bibinfo{author}{\bibfnamefont{R.~L.} \bibnamefont{Jaffe}},
  \bibinfo{author}{\bibfnamefont{M.}~\bibnamefont{Kardar}}, \bibnamefont{and}
  \bibinfo{author}{\bibfnamefont{A.}~\bibnamefont{Scardicchio}},
  \bibinfo{journal}{Phys. Rev. Lett.} \textbf{\bibinfo{volume}{96}},
  \bibinfo{pages}{080403} (\bibinfo{year}{2006}).

\bibitem[{\citenamefont{Bordag}(2006)}]{Bordag:2006b}
\bibinfo{author}{\bibfnamefont{M.}~\bibnamefont{Bordag}},
  \bibinfo{journal}{Phys. Rev. D} \textbf{\bibinfo{volume}{73}},
  \bibinfo{eid}{125018} (\bibinfo{year}{2006}).

\bibitem[{\citenamefont{Bordag and Nikolaev}(2009)}]{Bordag:kx}
\bibinfo{author}{\bibfnamefont{M.}~\bibnamefont{Bordag}} \bibnamefont{and}
  \bibinfo{author}{\bibfnamefont{V.}~\bibnamefont{Nikolaev}}
  (\bibinfo{year}{2009}), \bibinfo{note}{preprint arXiv:0911.0146}.

\bibitem[{\citenamefont{Gies and Klingmuller}(2006)}]{Gies:2006b}
\bibinfo{author}{\bibfnamefont{H.}~\bibnamefont{Gies}} \bibnamefont{and}
  \bibinfo{author}{\bibfnamefont{K.}~\bibnamefont{Klingmuller}},
  \bibinfo{journal}{Phys. Rev. Lett.} \textbf{\bibinfo{volume}{96}},
  \bibinfo{pages}{220401} (\bibinfo{year}{2006}).

\bibitem[{\citenamefont{Lifshitz}(1956)}]{Lifshitz56}
\bibinfo{author}{\bibfnamefont{E.~M.} \bibnamefont{Lifshitz}},
  \bibinfo{journal}{Sov. Phys. JETP} \textbf{\bibinfo{volume}{2}},
  \bibinfo{pages}{73} (\bibinfo{year}{1956}).

\bibitem[{\citenamefont{Palik}(1985)}]{Palik:85}
\bibinfo{author}{\bibfnamefont{E.~D.} \bibnamefont{Palik}},
  \emph{\bibinfo{title}{Handbook of Optical Constants of Solids}}
  (\bibinfo{publisher}{Academic}, \bibinfo{address}{New York},
  \bibinfo{year}{1985}).

\end{thebibliography}

\end{document}